\documentclass[aps,prl,reprint]{revtex4-1}
\usepackage{graphicx,color}
\usepackage{mathrsfs}
\usepackage{subfigure}
\usepackage{epstopdf}

\newcommand{\be}{\begin{eqnarray}}
\newcommand{\ee}{\end{eqnarray}}
\begin{document}

\title{Exact quantum field mappings between different experiments on quantum gases}

\author{Etienne Wamba$^{1,2}$}
\author{Axel Pelster$^{1}$}
\author{James R. Anglin$^{1}$}

\affiliation{\mbox{$^{1}$State Research Center OPTIMAS and Fachbereich Physik,} \mbox{Technische Universit\"at Kaiserslautern,} \mbox{67663 Kaiserslautern, Germany}\\
\mbox{$^{2}$International Center for Theoretical Physics,} \mbox{34151 Trieste, Italy}}

\begin{abstract}Experiments on trapped quantum gases can probe challenging regimes of quantum many-body dynamics, where strong interactions or non-equilibrium states prevent exact solutions. Here we present an exact result which holds even when no exact solutions can be found: a class of spacetime mappings of different experiments onto each other, as long as the gas particles interact via two-body potentials which possess a scaling property that most real interactions do possess. Since our result is an identity relating second-quantized field operators in the Heisenberg picture of quantum mechanics, it is otherwise general; it applies to arbitrary measurements on any mixtures of Bose or Fermi gases, in arbitrary initial states. Practical applications of this mapping include perfect simulation of non-trivial experiments with other experiments which may be easier to perform.\end{abstract}

\date{\today}

\maketitle

Spacetime coordinate transformations have long been used to map different solvable theoretical problems onto each other. A transformation introduced in 1890 by Poincar\'e \cite{poincare} has for example been used by Kustaanheimo and Stiefel to map the three-dimensional Kepler problem onto the four-dimensional harmonic oscillator \cite{stiefel1}, and thereby improve the numerical stability of perturbative calculations in celestial mechanics \cite{stiefel2}. The same mapping works in quantum mechanics \cite{duru1,duru2}, along with many other spacetime mappings between analytically solvable quantum systems \cite{jackiw,inomata,pelster1,kleinert,grosche}, such as that between the one-dimensional harmonic oscillator and free particle \cite{inomata,pelster1,kleinert}. Spacetime mappings have also been constructed between Markov processes \cite{pelster2}.

In quantum many-body theory, exactly solvable problems are rare, but spacetime mappings have been used in special cases to obtain additional evolution solutions by mapping them onto known ones. For quantum gases with certain special forms of inter-particle interaction, such as a $1/r^{2}$ potential \cite{Benj} or a short-ranged interaction with infinite scattering length \cite{Castin}, or for systems confined effectively to two spatial dimensions \cite{Pitaevskii}, non-trivial time-dependent many-body wave functions can be found exactly by taking a simpler known wave function, and transforming its space and time co-ordinates in a certain way. Scaling solutions have been found for general initial states within the Gross-Pitaevskii mean field approximation for the evolution of dilute Bose-Einstein condensates, either in two dimensions, or in further hydrodynamic approximation \cite{CastinDum,Kagan}, or in one dimension with an introduced imaginary potential \cite{Theocharis}, or with only three-body interactions \cite{Wu}. A spacetime transformation closely related to these scaling solutions has also been used, in mean-field theory in one dimension, to map evolution in time-dependent harmonic traps onto evolution with no trap, but with time-dependent interactions \cite{Etienne1,Etienne2}. 

Here we show that allowing time-dependent two-body interactions lets us extend the mean-field spacetime mapping to full quantum field theory, whose description of real quantum gases is itself essentially exact. This makes our class of spacetime mappings valid, not just between a few specially solvable theoretical problems, but between real experiments --- even if neither of the mapped experiments can be theoretically solved: see Fig.~1. Our mappings apply, moreover, to any mixtures of Bose or Fermi gases, in any number of dimensions, and for a class of interactions that includes most experimentally relevant cases. No restriction on initial states is required, because the mappings relate time-dependent operators in the Heisenberg picture of quantum mechanics, in which all quantum states are time-independent \cite{Cohen-Tannoudji}.

\begin{figure}[b]
\centering
\includegraphics[width=3.0in]{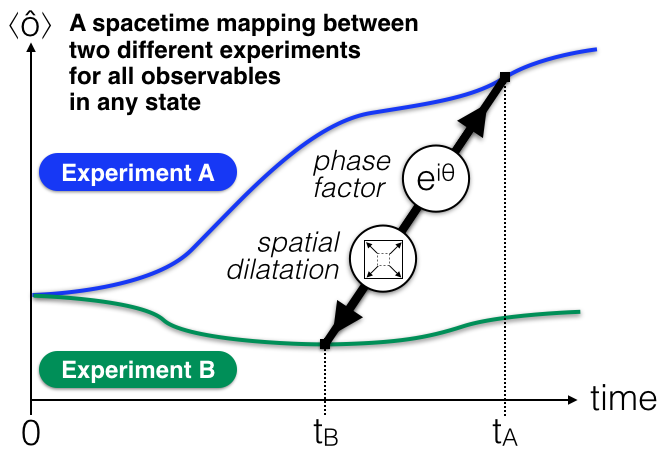}
\caption{Spacetime mapping between expectation values of an arbitrary observable $\hat{O}$ in different experiments. The mapping consists of a time-dependent dilatation of space and multiplication of field operators by a Gaussian phase factor, and it relates observables at different times in the two experiments, which may be very different procedures. A might for example be free expansion after turning off the trap, while B is ramping to a Feshbach resonance. Observables are the same in the two experiments initially, because the gas is prepared in the same (arbitrary) state. }\label{fig0}
\end{figure}

In the Heisenberg picture, the evolution of all observables is given, for any pure or mixed quantum state, by the equations of motion for the associated operators \cite{Cohen-Tannoudji}. For a quantum gas, all observables may be expressed in terms of the second-quantized field operator $\hat{\psi}_{n}(\mathbf{r},t)$, which destroys a particle of type $n$, and of its Hermitian conjugate field $\hat{\psi}^{\dagger}_{n}(\mathbf{r},t)$, which correspondingly creates a particle, at position $\mathbf{r}$ and time $t$. Since effectively one- or two-dimensional systems can be realized with ultracold atomic gases (by applying strong confining forces in transverse directions), we consider $\mathbf{r}$ to be in $D=1,2,$ or 3 dimensions. With the canonical (anti-)commutation relations
\begin{eqnarray}\label{CCR}
	[\hat{\psi}_{m}(\mathbf{r},t),\hat{\psi}^{\dagger}_{n}(\mathbf{r}',t)]_{\pm}=\delta_{mn}\delta^{D}(\mathbf{r}-\mathbf{r}')\;,
\end{eqnarray}
where $[\hat{A},\hat{B}]_{\pm}=\hat{A}\hat{B}\pm\hat{B}\hat{A}$, this description is equally applicable to fermions and bosons. Any experimental measurements can be expressed in terms of expectation values of $N$-point functions,
\begin{eqnarray}\label{Npoint}
F_{\mathbf{n},\mathbf{m}}(\mathbf{R},\mathbf{R}',t)&=&\left\langle \left[\Pi_{j=1}^{N}\hat{\psi}_{n_{j}}^{\dagger}(\mathbf{r}'_{j},t)\right]\left[\Pi_{j=1}^{N}\hat{\psi}_{m_{j}}(\mathbf{r}_{j},t)\right]\right\rangle\nonumber\\
\hbox{where }\mathbf{n}&=&\{n_{1},...,n_{N}\}\hbox{ and } \mathbf{R}=\{\mathbf{r}_{1},...,\mathbf{r}_{N}\}\;.
\end{eqnarray}
The time dependence of all observables is thus determined by the time dependence of the quantum fields. For a quantum gas whose particles may be of several species $n$ with possibly different masses $M_{n}$, with general two-body interactions in $D$ dimensions, the Heisenberg equation of motion for the field operator reads\begin{widetext}
\begin{eqnarray}\label{HE}
i\hbar\frac{\partial}{\partial t}\hat{\psi}_{n}(\mathbf{r},t) = \left[-\frac{\hbar^{2}\nabla^{2}}{2M_{n}} + V_{n}(\mathbf{r},t)\right]\hat{\psi}_{n}(\mathbf{r},t) + \sum_{klm}\int\!d^{D}r'\,U_{klmn}(\mathbf{r},\mathbf{r}',t)\hat{\psi}^{\dagger}_{k}(\mathbf{r}',t)\hat{\psi}_{l}(\mathbf{r}',t)\hat{\psi}_{m}(\mathbf{r},t)\;,
\end{eqnarray}
\end{widetext}
where $V_{n}(\mathbf{r},t)$ is the external potential felt by the particles of type $n$, and $U_{klmn}(\mathbf{r},\mathbf{r}',t)$ is the general two-particle interaction, which may possibly mix different particle species (such as by including spin flips) but cannot change particle masses (\textit{i.e.} $U_{klmn}=0$ except for $k,l,m,n$ such that $M_{k}=M_{l}$ and $M_{m}=M_{n}$). By exploiting collisional resonances controlled with time-dependent external fields \cite{Feshbach}, $U$ can also be made time-dependent in essentially any way, including being ramped to very large positive or negative values. 

Here we will consider cases where $U$ is a homogeneous function of its spatial arguments: $U_{klmn}(\lambda\mathbf{r},\lambda\mathbf{r}',t)=\lambda^{-s}U_{klmn}(\mathbf{r},\mathbf{r}',t)$, for some real number $s$, for any real factor $\lambda$. Most physically relevant interactions have this property, for some $s$; for a so-called contact interaction (Fermi-Huang pseudo-potential), $s=D$; for an electric or magnetic dipole-dipole interaction, $s=3$ (since experiments do not confine the electromagnetic fields into lower dimensions).

The spacetime mapping identity is as follows. Suppose that some particular set of time-dependent quantum fields $\hat{\psi}_{n}(\mathbf{r},t)$ satisfy (\ref{HE}), for some particular $V_{n}(\mathbf{r},t)$ and $U$. We then define a second set of quantum fields:
\begin{eqnarray}\label{dual}
\hat{\Psi}_{n}(\mathbf{r},t) = e^{-\frac{iM_{n}}{2\hbar}\frac{\dot{\lambda}}{\lambda}r^{2}}\lambda^{D/2}\hat{\psi}_{n}(\lambda \mathbf{r},\tau(t))
\end{eqnarray}
where $\lambda = \lambda(t)$, $\dot{\lambda}(t)\equiv d\lambda/dt$, and (importantly) $d\tau/dt = \lambda^{2}$.  The canonical (anti-)commutation relations (\ref{CCR}) for $\hat{\psi}_{n},\hat{\psi}_{n}^{\dagger}$ then imply that the $\hat{\Psi}_{n},\hat{\Psi}_{n}^{\dagger}$ satisfy the same relations and are just as canonical. 

Using the fact that $\hat{\psi}_{n}(\mathbf{r},t)$ obeys (\ref{HE}), it is then straightforward to show (see our Supplementary Material) that $\hat{\Psi}_{n}(\mathbf{r},t)$ also satisfies (\ref{HE}), but with $V_{n}\to\tilde{V}_{n}$ and $U\to\tilde{U}$, for
\begin{eqnarray}\label{tildevars}
\tilde{V}_{n}(\mathbf{r},t) &=& \lambda^{2}V_{n}(\lambda\mathbf{r},\tau(t))+\frac{M_{n}r^{2}}{2}\lambda^{3}\left(\frac{1}{\lambda^{2}}\frac{d}{dt}\right)^{2}\lambda\nonumber\\
\tilde{U}_{klmn}(\mathbf{r},\mathbf{r}',t) &=& [\lambda(t)]^{(2-s)}U_{klmn}(\mathbf{r},\mathbf{r}',t)\;.
\end{eqnarray}

This formal identity gains a concrete physical meaning when we further stipulate that $\lambda(0)=1$, $\dot{\lambda}(0)=0$, and $\tau(0)=0$, so that $\hat{\Psi}_{n}(\mathbf{r},0)=\hat{\psi}_{n}(\mathbf{r},0)$. At time $t=0$, therefore, the expectation values of any combination of $\hat{\Psi}_{n}$ and $\hat{\Psi}^{\dagger}_{n}$ operators, in any pure or mixed quantum state, will be identical to the expectation values, in the same quantum state, of the same combination of $\hat{{\psi}}_{n}$ and $\hat{{\psi}}^{\dagger}_{n}$ operators. The time-dependent $\hat{\Psi}_{n}$ and $\hat{{\psi}}_{n}$ operators therefore represent two different time evolutions of a quantum gas from the same initial conditions at $t=0$. 

By comparing the two different Heisenberg equations which they obey, we can see that $\hat{\psi}_{n}$ represents the gas evolving with $V_{n}$ and $U$, while $\hat{\Psi}_{n}$ represents the gas evolving with $\tilde{V}_{n}$ and $\tilde{U}$. The two evolutions which the mapping relates thus represent the same gas evolving under different experimental procedures.  Each of these two evolutions of an interacting quantum gas may be very complicated --- perhaps impossible to compute theoretically --- especially if the initial state is far from equilibrium; and the mapping is valid for any initial state.

To show what this means, we focus on a concrete example, in which $U$ is time-independent and $V_{n}=0$, but we achieve a constant, isotropic, harmonic potential $\tilde{V}_{n}=M_{n}\omega^{2}r^{2}/2$, having the same frequency $\omega$ for all species $n$, by choosing $\tau(t)=\omega^{-1}\tan(\omega t)$ and $\lambda(t) = \sec(\omega t)$  \cite{jackiwnote}. This indeed satisfies $\dot{\tau}=\lambda^{2}$, $\tau(0)=0$, $\lambda(0)=1$, and $\dot{\lambda}(0)=0$, but it provides $\tau(\frac{\pi}{2\omega})=\infty$. Hence an infinitely long time evolution of the $\hat{\psi}_{n}$ (for which $V_{n}=0$) is mapped onto the evolution of $\hat{\Psi}_{n}$ over only one quarter of a period of the harmonic trap with frequency $\omega$.

Furthermore, the time-independent $U$ has been mapped onto 
\begin{eqnarray}\label{}
	\tilde{U}_{klmn}(\mathbf{r},\mathbf{r}',t)= [\cos(\omega t)]^{s-2}U_{klmn}(\mathbf{r},\mathbf{r}')\;.
\end{eqnarray}
This is experimentally achievable, even though (depending on the sign of $s-2$), $[\cos(\omega t)]^{s-2}$ may approach either $\infty$ or 0 as $\omega t\to\pi/2$. For a contact interaction ($s=D$), for example, this can be achieved experimentally in $D=2$ by doing nothing, or with a time-dependent magnetic field which approaches either a Feshbach resonance \cite{Feshbach} (for $D=1$), or a point of zero scattering length between two Feshbach resonances \cite{Feshbach} (for $D=3$). The result then is that we have mapped the evolution of a gas with time-independent interactions, and no trap, onto the evolution of a gas in a time-independent harmonic trap, with a certain time-dependent interaction. 

The mapping is valid for any initial state, and how this state is prepared is of no theoretical consequence; for experimental convenience we can consider that the two experiments prepare their gases initially in the same isotropic harmonic trap, having the same frequency $\omega$ for all species, and with time-independent interactions $U$. In the first experiment (A), one simply turns off the trap at $t=0$, allowing the gas to expand until some final time $t_{\rm A}$. One then measures some $N$-point function (as in (\ref{Npoint})),
\begin{eqnarray}\label{Npoint2}
F^{\rm A}_{\mathbf{n},\mathbf{m}}(\mathbf{R},\mathbf{R}')&=&
\left\langle \left[\Pi_{j=1}^{N}\hat{\psi}_{n_{j}}^{\dagger}(\mathbf{r}'_{j},t_{\rm A})\right]\left[\Pi_{j=1}^{N}\hat{\psi}_{m_{j}}(\mathbf{r}_{j},t_{\rm A})\right]\right\rangle\;.\nonumber
\end{eqnarray}

In the second experiment (B), the trap is left on, but at $t=0$ one begins ramping a control parameter in such a way that $U\to [\cos(\omega t)]^{s-2}U=\tilde{U}$. One continues ramping until the final time $t_{\rm B}=\omega^{-1}\tan^{-1}(\omega t_{\rm A})<\pi/(2\omega)$. One then measures the $N$-point function, as in the A experiment. Since $\hat{\Psi}_{n}(\mathbf{r},t)$ is the solution to the Heisenberg equations of motion under the B experimental conditions, the $N$-point function at $t_{\rm B}$ will be
\begin{eqnarray}\label{Nprime}
F^{\rm B}_{\mathbf{n},\mathbf{m}}(\mathbf{R},\mathbf{R}')=\left\langle \left[\Pi_{j=1}^{N}\hat{\Psi}_{n_{j}}^{\dagger}(\mathbf{r}'_{j},t_{\rm B})\right]\left[\Pi_{j=1}^{N}\hat{\Psi}_{m_{j}}(\mathbf{r}_{j},t_{\rm B})\right]\right\rangle .\nonumber
\end{eqnarray}

Applying (\ref{dual}), however, and using the fact that $t_{\rm B}$ was defined by $\tan(\omega t_{\rm B})\equiv \omega t_{\rm A}$, we find the identity\begin{widetext}
\begin{eqnarray}\label{identity}
F^{\rm B}_{\mathbf{n},\mathbf{m}}(\mathbf{R},\mathbf{R}')&=&\frac{e^{-\frac{i\omega\,\tan(\omega t_{\rm B})}{2\hbar}\sum_{j=1}^{N}(M_{n_{j}}r_{j}^{2}-M_{m_{j}}r_{j}^{'2})}}{[\cos(\omega t_{\rm B})]^{ND}}
F^{\rm A}_{\mathbf{n},\mathbf{m}}\left(\frac{\mathbf{R}}{\cos(\omega t_{\rm B})},\frac{\mathbf{R}'}{\cos(\omega t_{\rm B})}\right)\;.
\end{eqnarray}
\end{widetext}
The identity (\ref{identity}) is an example of how our general spacetime mapping (\ref{dual}) implies concrete consequences: (\ref{identity}) explicitly relates arbitrary measurements on interacting quantum gases which evolve under significantly different experimental conditions, after being prepared in the same arbitrary initial state. The usefulness of this result is admittedly limited by the fact that the B experiment, with the time-dependent interaction in the constant trap, can only run for the maximum duration of a quarter trap-period. A lot can happen during this time, however, especially if the initial state is far from equilibrium --- and the mapping is valid for arbitrary states. It is also valid for arbitrary mixtures of Bose and/or Fermi gases, in arbitrarily many effective dimensions, having any two-particle interaction which is a homogeneous function of its spatial arguments (and which can be given the required time dependence in the B experiment). In the case of a contact interaction in one dimension ($s=1$), the interaction strength approaches infinity in the B experiment as $t_{\rm B}\to \pi/(2\omega)$, so it is possible to probe quite non-trivial many-body dynamics within our example scheme.

\begin{figure}[h!]
\centering
\includegraphics[width=3.0in]{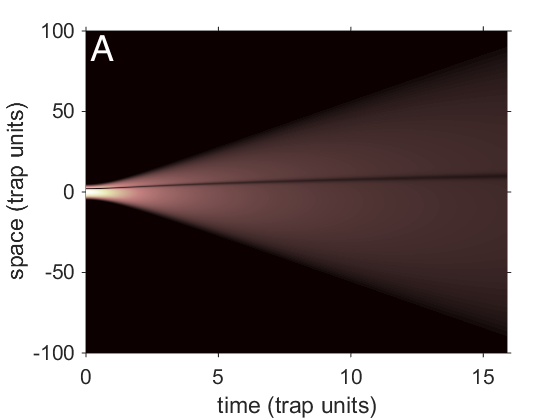}
\includegraphics[width=3.0in]{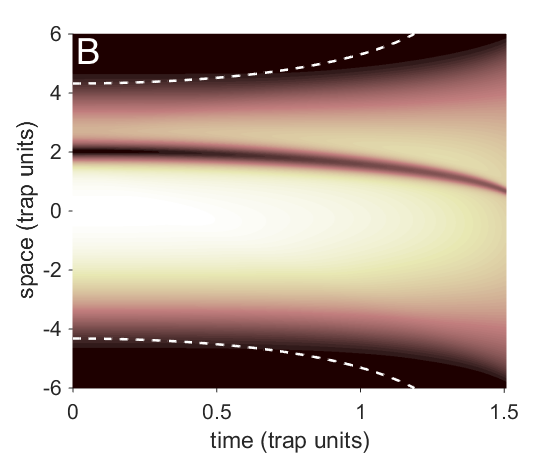}
\caption{Numerical results for Gross-Pitaevski mean-field evolution showing densities $|\psi(x,t)|^{2}$ (A) and $|\Psi(x,t)|^{2}$ (B), evolving under Eqn.~(\ref{HE}) with the operator fields replaced by complex classical fields, and a repulsive contact interaction whose initial strength is $U(x,x',0)=g(0)\delta(x-x')$ for $g(0)=50\hbar\omega a_{0}$ where $\omega$ is the initial trap frequency and $a_0=\sqrt{\hbar/(M\omega)}$ is the corresponding ground state width. The integrated densities are normalized to 1. Time and space are shown in units of $1/\omega$ and $a_0$, respectively; note the different ranges of space and time covered by the two plots. The two plots A and B correspond respectively to Experiments A (expansion with $V_{n}=0$, $g(t)=g(0)$) and B (ramped interaction with constant trap, $g(t)=g(0)\sec(\omega t)$), as described in the text. The spacetime transformation maps the two plots onto each other. Even when mean-field theory is not valid, the quantum field mapping remains exact; this Figure illustrates how it can relate non-trivial experiments. Here the initial state contains a dark soliton, which moves and changes in width while the whole cloud expands, and demonstrates how adiabaticity can break down at different times on different length scales. The dashed white curves in the B plot show the adiabatic Thomas-Fermi radius $R(t)=R(0)[\cos(\omega t)]^{-1/3}$. }\label{figs}
\end{figure}
We illustrate our mapping for that one-dimensional case in Fig.~2. We cannot plot a quantum field, but it is straightforward to show that the spacetime transformation (\ref{dual}) also serves to map between the classical field equations that are obtained by replacing the operator fields with complex c-number fields. These classical field theories are only mean-field approximations to quantum many-body dynamics, but they obey the same spacetime mapping identity, and thus serve to illustrate it.

Fig.~\ref{figs} shows $|\psi(x,t)|^{2}$ and $|\Psi(x,t)|^{2}$ corresponding to the Gross-Pitaevskii mean-field approximation to the gas density in experiments A and B, respectively, for a quasi-one-dimensional single-component Bose-Einstein condensate with a repulsive contact interaction $U(x,x',t)= g(t)\delta(x-x')$ whose initial state features a single dark soliton, slightly displaced from the center of the trap. Experiment A is the familiar scenario of free expansion with $g(t)=g(0)$. In experiment B, however, we see non-equilibrium response of the trapped gas to a temporally nonlinear ramping of the interaction strength $g(t)=g(0)\sec(\omega t)$. As the dashed lines in Fig.~2B show, the gas cloud at first expands adiabatically to follow the instantaneous Thomas-Fermi ground state, but when the equilibrium Thomas-Fermi radius increases too rapidly, the actual expansion of the gas fails to keep up. With increasingly strong nonlinearity, the soliton also narrows, but eventually it fails to shrink fast enough to maintain an equilibrium shape. We show in the Supplementary Material that the soliton width tracks its equilibrium value for a longer time than the Thomas-Fermi radius follows its equilibrium value, indicating that the gas loses equilibrium globally before losing it locally. 

Even just in mean-field theory, therefore, experiment B is non-trivial; and yet it is related exactly to free expansion by the spacetime mapping (\ref{dual}). Of course, the mean-field initial state is only accurate for a gas whose thermal excitation and quantum depletion are both negligible; and mean-field theory will eventually break down in any case, in both experiments, as the one-dimensional density becomes low (in A) or the interaction becomes strong (in B). The validity of our mapping for the quantum field operators, however, means that whatever the actual quantum evolution of the gas may be, the observations in A and B scenarios will still be related by (\ref{identity}).

Time-dependent potentials and interactions are well-established experimental tools in today's quantum gas labs, and the mapping between an isotropic harmonic trap and no trap was just one special case of ${V}_{n}$ and $\lambda(t)$. Since our exact spacetime mapping is so general, allowing arbitrary $\lambda(t)$ and applying to arbitrary measurements on arbitrary mixtures of multi-component Bose and Fermi gases with many realistically possible interactions, prepared in any initial states, it is a strong prediction from quantum field theory which can be tested in a wide range of real quantum gas systems. If the mapping is experimentally confirmed, it can become a tool to expand experimental technique, by allowing time-dependent traps to mimic time-dependent interactions, or vice versa; or it may be used to test for experimental errors. The mapping identity may also be a useful benchmark for theoretical approximations: failure to fulfill it will mark limits of validity.

\begin{acknowledgements}
\section{Acknowledgements}
EW acknowledges funding from the Alexander von Humboldt Foundation, and from the Abdus Salam International Center for Theoretical Physics, through a Simons Associateship. AP thanks the German Research Foundation (DFG) for support via the Collaborative Research Center SFB/TR 49 ``Condensed Matter Systems with Variable Many-Body Interactions''.
\end{acknowledgements}
\clearpage
\setcounter{equation}{0}
\setcounter{figure}{0}
 \renewcommand{\theequation}{S-\arabic{equation}}
 \renewcommand{\thefigure}{S-\arabic{figure}}

\begin{widetext}\begin{center}\textbf{Supplementary Material}\end{center}

\textbf{Derivation of the mapping identity}

The canonical (anti-)commutation relations (1) for $\hat{\psi}_{n}$ and $\hat{\psi}_{n}^{\dagger}$ imply that the $\hat{\Psi}_{n}$ and $\hat{\Psi}_{n}^{\dagger}$ satisfy precisely the same relations:
\begin{eqnarray}\label{CCR2}
	[\hat{\Psi}_{m}(\mathbf{r},t),\hat{\Psi}^{\dagger}_{n}(\mathbf{r}',t)]_{\pm}&=&\delta_{mn}\lambda^{D}\delta^{D}\Big(\lambda(\mathbf{r}-\mathbf{r}')\Big)\equiv\delta_{mn}\delta^{D}(\mathbf{r}-\mathbf{r}')\;.
\end{eqnarray}

To show that $\hat{\Psi}_{n}$ also satisfy (3), but with $V_{m}\to\tilde{V}_{n}$ and $U\to\tilde{U}$, we first differentiate (4) with respect to $t$ and find
\begin{eqnarray}\label{}
	i\hbar\frac{\partial}{\partial t}\hat{\Psi}_{n}(\mathbf{r},t) = e^{-\frac{iM_{n}}{2\hbar}\frac{\dot{\lambda}}{\lambda}r^{2}}\lambda^{D/2}\left[i\hbar \frac{D\dot{\lambda}}{2\lambda} +\frac{M_{n}r^{2}}{2}\left(\frac{\ddot{\lambda}}{\lambda}-\frac{\dot{\lambda}^{2}}{\lambda^{2}}\right)
+i\hbar \frac{\dot{\lambda}}{\lambda}\mathbf{r}\cdot\mathbf{\nabla}+i\hbar\frac{\partial \tau}{\partial t}\frac{\partial}{\partial \tau}\right]\hat{\psi}_{n}\big(\lambda\mathbf{r},\tau(t)\big)\;,
\end{eqnarray}
where the partial differentiation with respect to $\tau$ implies treating $\lambda$ as a constant --- the differentiation of $\lambda(t)\mathbf{r}$ with respect to $t$, in the argument of $\hat{\psi}$, is the preceding $\mathbf{r}\cdot\mathbf{\nabla}$ term.
Using the Heisenberg equation (3) for $\partial_{\tau}\hat{\psi}_{n}(\lambda\mathbf{r},\tau)$ 
\begin{eqnarray}\label{}
i\hbar\frac{\partial}{\partial\tau}\hat{\psi}_{n}(\lambda\mathbf{r},\tau)&=&\left[-\frac{\hbar^{2}}{2M_{n}}\nabla_{\lambda\mathbf{r}}^{2}+V_{n}(\lambda \mathbf{r},\tau)\right]\hat{\psi}_{n}(\lambda\mathbf{r},\tau)\nonumber\\
&&+ \sum_{klm}\int\!d^{D}(\lambda r')\,U_{klmn}(\lambda\mathbf{r},\lambda\mathbf{r}',\tau)\hat{\psi}^{\dagger}_{k}(\lambda\mathbf{r}',\tau)\hat{\psi}_{l}(\lambda\mathbf{r}',\tau)\hat{\psi}_{m}(\lambda\mathbf{r},\tau)
\end{eqnarray}
where the Laplacian with respect to $\lambda\mathbf{r}$ is simply 
\begin{equation}
\nabla_{\lambda\mathbf{r}}^{2}\equiv  \sum_{j=1}^{D}\frac{\partial^{2}}{\partial (\lambda x_{j})^{2}} \equiv \lambda^{-2}\nabla^{2}\;,\nonumber
\end{equation}
and also using the ``homogeneous function'' property of the interaction $U_{klmn}(\lambda\mathbf{r},\lambda\mathbf{r}',\tau)=\lambda^{-s}U_{klmn}(\mathbf{r},\mathbf{r}',\tau)$, this becomes
\begin{eqnarray}\label{}
	i\hbar\frac{\partial}{\partial t}\hat{\Psi}_{n}(\mathbf{r},t)&=& e^{-\frac{iM_{n}}{2\hbar}\frac{\dot{\lambda}}{\lambda}r^{2}}\lambda^{D/2}\left[i\hbar\frac{D\dot{\lambda}}{2\lambda} +\frac{M_{n}r^{2}}{2}\left(\frac{\dot{\lambda}}{\lambda}\right)^{2}+i\hbar\frac{\dot{\lambda}}{\lambda}\mathbf{r}\cdot\mathbf{\nabla}-\frac{d\tau}{dt}\frac{1}{\lambda^{2}}\frac{{\hbar}^{2}}{2M_{n}}\nabla^{2}\right]\hat{\psi}_{n}(\lambda\mathbf{r},\tau)\nonumber\\
	&&+	e^{-\frac{iM_{n}}{2\hbar}\frac{\dot{\lambda}}{\lambda}r^{2}}\lambda^{D/2}\left[\frac{d\tau}{dt}V_{n}(\lambda\mathbf{r},\tau)+\frac{M_{n}r^{2}}{2}\left(\frac{\ddot{\lambda}}{\lambda}-2\frac{\dot{\lambda}^{2}}{\lambda^{2}}\right)\right]\hat{\psi}_{n}(\lambda\mathbf{r},\tau)
	\nonumber\\
	&&+e^{-\frac{iM_{n}}{2\hbar}\frac{\dot{\lambda}}{\lambda}r^{2}}\lambda^{D/2}\frac{d\tau}{dt}\lambda^{D-s}\sum_{klm}\int\!d^{D}r'\,U_{klmn}(\mathbf{r},\mathbf{r}',\tau)\hat{\psi}^{\dagger}_{k}(\lambda\mathbf{r}',\tau)\hat{\psi}_{l}(\lambda\mathbf{r}',\tau)\hat{\psi}_{m}(\lambda\mathbf{r},\tau)\;,
	\end{eqnarray}
where on the right-hand side $\tau$ denotes $\tau(t)$ everywhere.	
Then using the definition $d\tau/dt=\lambda^{2}$ this yields
\begin{eqnarray}\label{}
	i\hbar\frac{\partial}{\partial t}\hat{\Psi}_{n}(\mathbf{r},t)&=& e^{-\frac{iM_{n}}{2\hbar}\frac{\dot{\lambda}}{\lambda}r^{2}}\lambda^{D/2}
			\left[i\hbar\frac{D\dot{\lambda}}{2\lambda} +\frac{M_{n}r^{2}}{2}\left(\frac{\dot{\lambda}}{\lambda}\right)^{2}+i\hbar\frac{\dot{\lambda}}{\lambda}\mathbf{r}\cdot\mathbf{\nabla}
							-\frac{{\hbar}^{2}}{2M_{n}}\nabla^{2}\right]\hat{\psi}_{n}(\lambda\mathbf{r},\tau)\nonumber\\
	&&+	e^{-\frac{iM}{2\hbar}\frac{\dot{\lambda}}{\lambda}r^{2}}\lambda^{D/2}
	\left[\lambda^{2}V_{n}(\lambda\mathbf{r},\tau)+\frac{M_{n}r^{2}}{2}\lambda^{3}\left(\frac{1}{\lambda^{2}}\frac{d}{dt}\right)^{2}\lambda\right]
	\hat{\psi}_{n}(\lambda\mathbf{r},\tau)\nonumber\\
	&&+e^{-\frac{iM_{n}}{2\hbar}\frac{\dot{\lambda}}{\lambda}r^{2}}\lambda^{3D/2}\lambda^{2-s}\sum_{klm}\int\!d^{D}r'\,U_{klmn}(\mathbf{r},\mathbf{r}',\tau)\hat{\psi}^{\dagger}_{k}(\lambda\mathbf{r}',\tau)\hat{\psi}_{l}(\lambda\mathbf{r}',\tau)\hat{\psi}_{m}(\lambda\mathbf{r},\tau)\\
	&\equiv&-\frac{\hbar^{2}}{2M_{n}}\nabla^{2}\left[e^{-\frac{iM_{n}}{2\hbar}\frac{\dot{\lambda}}{\lambda}r^{2}}\lambda^{D/2}\hat{\psi}_{n}\big(\lambda\mathbf{r},\tau(t)\big)\right]+\left[\lambda^{2}V_{n}\big(\lambda\mathbf{r},\tau(t)\big)+\frac{M_{n}r^{2}}{2}\lambda^{3}\left(\frac{1}{\lambda^{2}}\frac{d}{dt}\right)^{2}\lambda\right]\hat{\Psi}_{n}(\mathbf{r},t)\nonumber\\
	&&+\lambda^{2-s}\sum_{klm}\int\!d^{D}r'\,U_{klmn}\big(\mathbf{r},\mathbf{r}',\tau(t)\big)\hat{\Psi}^{\dagger}_{k}(\mathbf{r}',t)\hat{\Psi}_{l}(\mathbf{r}',t)\hat{\Psi}_{m}(\mathbf{r},t)\end{eqnarray}\end{widetext}
where in the last line we have used the mass conservation property of non-relativistic interactions ($U_{klmn}=0$ except for $k,l,m,n$ such that $M_{k}=M_{l}$ and $M_{m}=M_{n}$) to replace $\hat{\psi}_{j}\to\hat{\Psi}_{j}$ by inserting additional phase factors which all cancel each other.

Recognizing the first term in square brackets on the right-hand side of (S-7) as $\hat{\Psi}_{n}(\mathbf{r},t)$, we confirm the statements (5) in our main text.
\\
\\
\textbf{Mapped version of Fig.~2.}

\begin{figure}[tbh]
\centering
\includegraphics[width=3.0in]{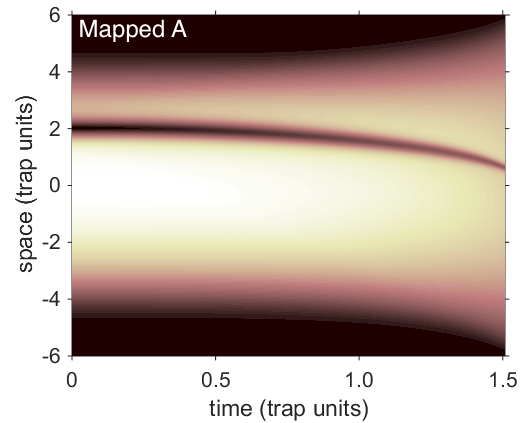}
\includegraphics[width=3.0in]{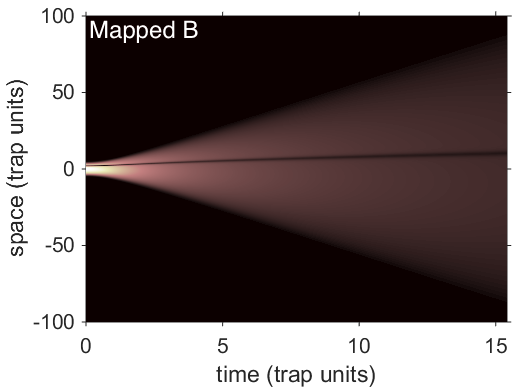}
\caption{Space-time evolution of the densities $|\psi(x,t)|^{2}$ and $|\Psi(x,t)|^{2}$, respectively, plotted by mapping the densities obtained in experiments A (above) and B (below), according to (4) and its inverse. Since the mapping is exact, the barely visible differences between these plots, and those of Fig.~2 in our main text, are due to large re-scalings of numerical solutions with finite resolution.}\label{mfigs}
\end{figure}

Fig.~\ref{mfigs} shows $|\psi(x,t)|^{2}$ and $|\Psi(x,t)|^{2}$ obtained by mapping the Gross-Pitaevskii mean fields from the evolutions shown in Fig.~2. The  grids of numbers shown in each plot have been transformed according to (4) and its inverse, but it is effectively impossible to tell that the two plots of Fig.~2 have not just been switched. The mapping is perfect: free expansion over about $15/(2\pi)$ trap periods and 160 trap widths corresponds exactly to interaction ramping (with the particular $1/\cos(\omega t)$ time dependence specified in the main text) over about $1.5/(2\pi)$ trap periods and 10 trap widths. Remember, however, that the mapping is not trivial: the spatial dilatation factor $\lambda$ is time-dependent, and the time transformation $t\to\tau(t)$ is nonlinear.
\\
\\
\textbf{Adiabaticity on different length scales in Experiment B}

Figure 2B in our main text has shown the mean-field approximation to the time-dependent condensate density during Experiment B, in which the strength of the repulsive contact interaction is ramped towards infinity. The white dashed curves superimposed on the density plot show the adiabatic Thomas-Fermi radius, which would mark the approximate edge of the condensate cloud, if the ramping of the interaction strength were infinitely slow. As observed in the main text, the condensate does expand when the repulsive interaction increases; but as the interaction increases more and more rapidly, the increasing pressure which it supplies cannot expand the gas fast enough to maintain the even more rapidly increasing equilibrium size. The actual condensate size therefore falls below the instantaneous equilibrium Thomas-Fermi radius (at least within Gross-Pitaevskii mean-field theory).

We can quantify this breakdown in adiabaticity of the condensate cloud size by fitting the actual condensate density profile to a Thomas-Fermi-like parabola, 
\begin{eqnarray}\label{parabola}
|\Psi(x,t)|^{2} \to \frac{M\omega^{2}}{2 g_{\rm TF}(t)}[R^{2}(t)-x^{2}],
\end{eqnarray}
tuning $R(t)$ and $g_{\rm TF}(t)$ independently at each instant $t$. These fits are quite good --- the actual density envelope remains quite parabolic, even though the width of the parabola lags behind the adiabatic Thomas-Fermi value. From this time-dependent parabolic fit, we obtain the fitted instantaneous interaction strength $g_{\rm TF}(t)$ for which that parabolic density profile would be the Thomas-Fermi ground state. The resulting $g_{\rm TF}(t)$ is shown as the dotted curve in Fig.~S-2, along with the actual interaction strength $g(t) = g(0)\sec(\omega t)$, shown as a solid curve. Since the initial state has been well relaxed (by imaginary time evolution with fixed normalization), at an initial interaction $g(0)$ large enough for the Thomas-Fermi approximation to Gross-Pitaevskii density envelope to be quite accurate, the initial fit value $g_{\rm TF}(0)$ coincides closely with the actual interaction strength $g(0)$. As the time-dependent condensate fails to sustain the rapidly increasing Thomas-Fermi radius of instantaneous equilibrium, the Thomas-Fermi-fitted effective interaction strength $g_{\rm TF}(t)$ fails to rise as fast as the actual $g(t)$:  the dashed curve falls away from the solid curve significantly from about half-way through the plotted time interval.

The initial state used for both A and B plots was prepared by relaxation (Gross-Pitaevskii evolution in imaginary time) while maintaining $\int dx\,|\psi|^{2}=1$, with interaction $g(0)=50\hbar\omega a_{0}$. By including a zero in the trial wave function before relaxation, however, and stopping relaxation before the zero filled in, the initial state was prepared with a dark soliton at $x=2$. In the subsequent real time evolution shown in the plots, this soliton moves and changes width, while the whole cloud expands; in Experiment B, the soliton narrows as the repulsive interaction strengthens. Such a narrowing would be expected adiabatically, if the interaction were increased very slowly; when the interaction strengthens too quickly, however, the soliton may not have time to become as narrow as it would in the adiabatic limit.  

The adiabatic soliton width is determined by three factors: the `grayness' of the soliton (which increases when the soliton moves); the local gas density near the soliton, which is well described by the instantaneous Thomas-Fermi envelope described above; and the interaction strength $g$. To quantify how closely the soliton narrowing adapts to the changing interaction strength, independently of the soliton grayness and ambient density, we can approximate the harmonic potential as constant over the width of the soliton, and then approximate the Gross-Pitaevskii evolution of $\Psi(x,t)$ near the soliton with the integrable nonlinear Schr\"odinger equation
\begin{equation}\label{GPAD}
i\hbar\partial_{t}\tilde{\Psi}(x,t) = \left[-\frac{\hbar^{2}}{2M}\partial_{xx} + V_{0} + g_{\rm sol}|\tilde{\Psi}|^{2}\right]\tilde{\Psi}(x,t)\;
\end{equation}
where the effective trap potential at the soliton $V_{0}$, and the effective interaction $g_{\rm sol}$ which is `felt' by the soliton, are considered as constants. Moving grey soliton solutions to (\ref{GPAD}) are given by the ansatz
\begin{widetext}
\begin{eqnarray}\label{ansatz}
\tilde{\Psi}(x,t) = \left(i\beta + \kappa\tanh\left[\kappa\frac{\sqrt{g_{\rm sol}M}}{\hbar}\left(x-x_{0}-\beta\frac{\sqrt{g_{\rm sol}}}{\sqrt{M}}t\right)\right]\right)\exp\left(-\frac{i}{\hbar}[V_{0}+g_{\rm sol}(\beta^{2}+\kappa^{2})]t\right)
\end{eqnarray}
\end{widetext}
for any constants $\beta$, $\kappa$, and $x_{0}$.
 
\begin{figure}[tbh]
\centering
\includegraphics[width=3.0in]{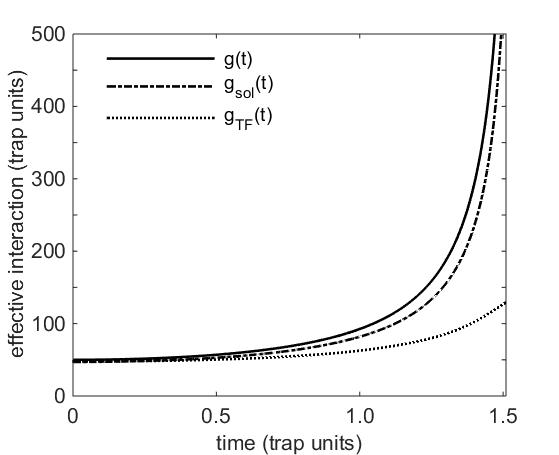}
\caption{Time dependence in trap units $\hbar\omega a_{0}$ of the actual contact interaction strength $g(t)=g(0)\sec(\omega t)$ (solid curve) and of two effective interaction strengths $g_{\rm sol}(t)$ (dash-dotted curve) and $g_{\rm TF}(t)$ (dotted curve), obtained by fitting small- and large-scale features of the density profile, respectively, as explained in the Supplementary text. The effective interaction $g_{\rm TF}(t)$ represents the constant interaction strength for which the instantaneous parabolic envelope of the density profile would be a Thomas-Fermi ground state of the form (\ref{parabola}). The effective interaction $g_{\rm sol}(t)$ represents the constant interaction strength for which the part of $\Psi(x,t)$ near the soliton would be a gray soliton solution to the integrable nonlinear Schr\"odinger equation obtained by approximating the harmonic potential as constant. The fact that $g_{\rm sol}(t)$ follows $g(t)$ more closely than $g_{\rm TF}(t)$ shows that the condensate remains locally adiabatic on short length scales for longer than it remains adiabatic globally.}
\end{figure}

For each time $t$, then, we fit the actual density profile $|\Psi(x,t)|^{2}$ to the ansatz $|\tilde{\Psi}(x,t)|^{2}$ near the soliton, by tuning $\kappa$, $\beta$, $x_{0}$ and $g_{\rm sol}$. The fits remain good, near the soliton, at all times. The resulting fitted $g_{\rm sol}(t)$ then represents the effective interaction strength `felt' by the soliton at time $t$: it is the constant interaction strength for which the instantaneous $\Psi(x,t)$ would be a gray soliton solution locally, given the instantaneous grayness parameter $\beta$ and local envelope density $\kappa^{2}+\beta^{2}$. This fitted $g_{\rm sol}(t)$ is shown as the dash-dotted curve in Fig.~S-2. Like the analogous effective interaction strength $g_{\rm TF}(t)$ implied by the overall condensate radius, the effective interaction strength $g_{\rm sol}(t)$ implied by the instantaneous soliton width also lags behind the actual interaction strength $g(t)$ as it increases. (It seems to follow a power law: $g_{\rm sol}(t)/g(0)\doteq [g(t)/g(0)]^{0.89}$.) We can note, however, that $g_{\rm sol}(t)$ follows the actual $g(t)$ much more closely than $g_{\rm TF}(t)$ does (which does not appear to obey any comparable power law). The soliton width behaves more adiabatically, for longer, than the overall condensate cloud size.

This differential adiabaticity, depending on length scale, is physically intuitive. The dynamical time scale for large-scale changes of the overall density profile is that of low-frequency collective modes of the condensate: it is on the order of the harmonic trap period. The soliton, however, is a structure on the scale of the local healing length; the characteristic time scale for evolution on this shorter length scale is correspondingly shorter. We therefore expect that the more rapidly responding soliton width will be better able to follow the changing $g(t)$ than the more slowly reacting Thomas-Fermi radius. Adiabaticity breaks down on large scales (globally) before it breaks down on small scales (locally).

The evolution in Experiment B is therefore quite interesting, because for initial conditions which feature structures on different length scales, it can reveal how adiabaticity and equilibration occur at different rates on those different length scales, so that a quantum gas may be both in and out of equilibrium, in different respects, at the same time. The failure of mean-field theory at stronger interactions, or even initially because of quantum or thermal depletion, will invalidate our plotted mean-field evolution; but it will only make the real Experiment B even more interesting. This non-trivial experiment may nonetheless be simulated exactly in every respect, with all possible quantum and thermal and non-equilibrium corrections fully included, by the standard expansion of Experiment A --- when the exact spacetime mapping is applied.


\begin{thebibliography}{10}
\bibitem{jackiwnote}The transformation in our example happens to be the same transformation that was defined in Ref.~\cite{jackiw}.
\end{thebibliography}
\end{document}